\documentclass[12pt]{article} % For LaTeX2e
\usepackage[colorlinks, citecolor={blue}]{hyperref}
\usepackage{url}
\usepackage{amsfonts,amscd,amssymb}
\usepackage{amsthm,amsmath,natbib}
\usepackage{algorithm,algorithmicx,algpseudocode}
\usepackage{bm}
\usepackage{bbm} %bb font numbers
\usepackage[table]{xcolor}
\usepackage{verbatim}
\usepackage{graphicx}
\usepackage{setspace}
\usepackage{natbib}
\usepackage[margin=1in]{geometry}
\usepackage{enumitem}
\usepackage{listings}
\usepackage[textsize=tiny]{todonotes}
\usepackage{tikz}
\usepackage{etoolbox}
\usepackage{appendix}
\usepackage[format=plain,
font=it]{caption}
\usepackage{subcaption}
\usepackage{wrapfig}
\usepackage{xr}
\usepackage{booktabs}
\usepackage{multirow}
\usepackage{authblk}

\usetikzlibrary{matrix}
\usetikzlibrary{backgrounds}
\usetikzlibrary{calc}
\usetikzlibrary{arrows,shapes}
\usetikzlibrary{decorations}
\usetikzlibrary{decorations.pathmorphing}
\usetikzlibrary{fit}
\usetikzlibrary{decorations.pathreplacing}

\newtoggle{quickdraw}
\toggletrue{quickdraw} % Uncomment this to render more quickly (non-random)

\definecolor{lightgrey}{rgb}{0.9,0.9,0.9}
\definecolor{darkgreen}{rgb}{0,0.3,0}
\definecolorset{rgb}{}{}{darkred,0.8,0,0;darkgreen,0,0.5,0;darkblue,0,0,0.5}

\DeclareMathOperator*{\argmin}{arg\,min}

\newcommand{\density}[2][]{\ensuremath{\pi_{#1}(#2)}}

\definecolor{trevorblue}{rgb}{0.330, 0.484, 0.828}
\definecolor{trevoryellow}{rgb}{0.829, 0.680, 0.306}

\newcommand{\complexity}{\mathcal{O}}

\DeclareRobustCommand{\disambiguate}[3]{#2~#3}
\newcommand{\paramIndex}{p}
\newcommand{\given}{\, | \,}
\newcommand{\Var}{\textrm{Var}}
\newcommand{\transposeSymbol}{\text{\raisebox{.5ex}{$\intercal$}}}
\newcommand{\normal}{\text{Normal}}
\newcommand{\bernoulli}{\text{Bernoulli}}

\newcommand{\gscale}{\tau}
\newcommand{\lscale}{\lambda}
\newcommand{\blscale}{\bm{\lambda}}
\newcommand{\Lscale}{\Lambda}
\newcommand{\bLscale}{\bm{\Lscale}}
\newcommand{\regcoef}{\theta}
\newcommand{\bregcoef}{\bm{\regcoef}}
\newcommand{\bomega}{\bm{\omega}}
\newcommand{\bOmega}{\bm{\Omega}}
\newcommand{\Precision}{\bm{\Phi}}
\newcommand{\CholeskyFactor}{\bm{L}}
\newcommand{\localPrior}[1][\cdot]{\density[\rm local]{#1}}
\newcommand{\globalPrior}[1][\cdot]{\density[\rm global]{#1}}

\title{Computational Statistics and Data Science in the Twenty-first Century\footnote{Chapter 1 in Piegorsch, 
		W.W., Levine, R.A., Zhang, H.H., and Lee, T.C.M. (eds.). 2022. Computational 
		Statistics in Data Science. Chichester: 
		John Wiley \& Sons.}}
\date{}

\author[1]{Andrew J.~Holbrook}
\author[2]{Akihiko Nishimura}
\author[3]{Xiang Ji}
\author[1,4,5]{Marc A.~Suchard}
\affil[1]{Department of Biostatistics, University of California, Los Angeles}
\affil[2]{Department of Biostatistics, Johns Hopkins University}
\affil[3]{Department of Mathematics, Tulane University}
\affil[4]{Department of Biomathematics, University of California, Los Angeles}
\affil[5]{Department of Human Genetics, University of California, Los Angeles}

\graphicspath{{figures/}}

\begin{document}

\maketitle

\begin{abstract}

Data science has arrived, and computational statistics is its engine.  As the scale and complexity of scientific and industrial  data grow, the discipline of computational statistics  assumes an increasingly central role among the statistical sciences.  An explosion in the range of real-world applications means the development of more and more specialized computational methods, but five Core Challenges remain.  We provide a high-level introduction to computational statistics by focusing on its central challenges, present recent model-specific advances and preach the ever-increasing role of non-sequential computational paradigms such as multi-core, many-core and quantum computing.  Data science is bringing major changes to computational statistics, and these changes will shape the trajectory of the discipline in the 21st century.

\end{abstract}

\section{Introduction}

We are in the midst of the data science revolution. In October 2012, the Harvard Business Review famously declared data scientist the sexiest job of the 21st century \citep{davenport2012data}. By September 2019, Google searches for the term `data science'  had multiplied over 7-fold \citep{trends}, one multiplicative increase for each intervening year.  In the U.S.~between the years 2000 and 2018, the number of bachelor's degrees awarded in either statistics or biostatistics increased  over 10-fold (382 to 3,964) and the number of doctoral degrees almost trippled (249 to 688) \citep{ASA}.  In 2020, seemingly every major university has established or is establishing its own data science institute, center or initiative.

\emph{Data science} \citep{cleveland2001data,donoho201750} combines multiple pre-existing disciplines (e.g., statistics, machine learning, computer science) with a redirected  focus on creating, understanding and systematizing workflows that turn real-world data into actionable conclusions.
The ubiquity of data in all economic sectors and scientific disciplines makes data science eminently relevant to cohorts of researchers for whom the discipline of statistics was previously closed-off and esoteric.
Data science's emphasis on practical application only enhances the importance of \emph{computational statistics}, the interface between statistics and computer science primarily  concerned with the development of algorithms producing either statistical inference%
\footnote{\emph{Statistical inference} is an umbrella term for hypothesis testing, point estimation and the generation of (confidence or credible) intervals for population functionals (mean, median, correlations, etc.) or model parameters.} %
or predictions.
Since both of these products comprise essential tasks in any data scientific workflow, we believe that the pan-disciplinary nature of data science only increases the number of  opportunities for computational statistics to evolve by taking on new applications%
\footnote{We present the problem of phylogenetic reconstruction in Section \ref{sec:phylogen} as one such example arising from the field of molecular epidemiology.} %
and serving the needs of new groups of researchers.

This is the natural role for a discipline that has increased the breadth of statistical application from the beginning.
First put forward by R.A.~Fisher in 1936 \citep{fisher1960design,fisher1992statistical}, the permutation test allows the scientist (who owns a computer) to test hypotheses about a broader swath of functionals of a target population  while making fewer statistical assumptions \citep{wald1944statistical}.  With a computer, the scientist uses the bootstrap \citep{efron1992bootstrap,efron1994introduction} to obtain confidence intervals for population functionals and parameters of models too complex for analytic methods.  Newton-Raphson optimization and the Fisher scoring algorithm facilitate linear regression for binary, count and categorical outcomes \citep{bliss,mccullagh1989generalized}.   More recently, Markov chain Monte Carlo (MCMC) \citep{tierney1994markov,brooks2011handbook} has made Bayesian inference practical for massive, hierarchical and highly structured models that are useful for the analysis of a significantly wider range of scientific phenomena.

While computational statistics increases the diversity of statistical applications historically, certain central difficulties exist and will continue to remain for the rest of the 21st century.  In Section \ref{sec:core1}, we present the first class of Core Challenges, or challenges that are easily quantifiable for generic tasks. Core Challenge 1 is Big $N$, or statistical inference when the number `\emph{N}' of observations or data points is large.  Core Challenge 2 is Big $P$, or statistical inference when the model parameter count `\emph{P}' is large.  And Core Challenge 3 is Big $M$, or statistical inference when the model's objective or density function is multimodal (having many modes `$M$')\footnote{The use of `\emph{N}' and `\emph{P}' to denote observation and parameter count is common.  We have taken liberties in coining the use of `\emph{M}' to denote mode count.}.   When large, each of these quantities brings its own unique computational difficulty.  Since well over 2.5 exabytes (or $2.5\times10^{18}$ bytes) of data come into existence each day \citep{chavan2014survey}, we are confident Core Challenge 1 will survive well into the 22nd century.

But Core Challenges 2 and 3 will also endure: data complexity often increases with size, and researchers strive to understand increasingly complex phenomena. Because many examples of big data become `big' by combining heterogeneous sources, big data often necessitate big models.
With the help of two recent examples, Section \ref{sec:modSpecific} illustrates how computational statisticians make headway at the intersection of big data and big models with model-specific advances.
In Section \ref{sec:sparse}, we present recent work in Bayesian inference for big \emph{N}, big \emph{P} regression.
Beyond the simplified regression setting, data often come with structures (e.g., spatial, temporal, network), and correct inference must take these structures into account.  For this reason, we present novel computational methods for a highly structured and hierarchical model for the analysis of multi-structured, epidemiological data in Section \ref{sec:phylogen}.

The growth of model complexity leads to new inferential challenges.   Whereas we define Core Challenges 1-3 in terms of generic target distributions or objective functions, Core Challenge 4 arises from inherent difficulties in treating complex models generically.   Core Challenge 4 (Section \ref{sec:genalg}) describes the difficulties and trade-offs that must be overcome to create fast, flexible and friendly `algo-ware'.  This Core Challenge requires the development of statistical algorithms that maintain efficiency despite model structure and, thus, apply to a wider swath of target distributions or objective functions `out-of-the-box'.  Such generic algorithms typically require little cleverness or creativity to implement, limiting the amount of time data scientists must spend worrying about computational details.    Moreover, they aid the development of flexible statistical software that adapts to complex model structure in a way that users easily understand.
But it is not enough that software be flexible and easy to use: mapping computations to computer hardware for optimal implementations remains difficult.  In Section \ref{sec:hardware}, we argue that Core Challenge 5, effective use of computational resources such as central and graphics processing units (CPU and GPU) and quantum computers, will become increasingly central to the work of the computational statistician as data grow in magnitude.

\section{Core challenges 1-3}\label{sec:core1}

Before providing two recent examples of 21st century computational statistics (Section \ref{sec:modSpecific}), we present three easily quantified Core Challenges within computational statistics that we believe will always exist: big $N$, or inference from many observations; big $P$, or inference with high-dimensional models; and big $M$, or inference with non-convex objective---or multimodal density---functions.  In 21st century computational statistics, these challenges often co-occur, but we consider them separately in this section.

\subsection{Big \emph{N}}\label{sec:bigN}

Having a large number of observations makes different computational methods difficult in different ways.  A worst case scenario, the \emph{exact} permutation test requires the production of $N!$ datasets.  Cheaper alternatives, resampling methods such as the Monte Carlo permutation test or the boostrap, may require anywhere from thousands to hundreds of thousands of randomly produced datasets \citep{wald1944statistical,efron1994introduction}. When, say, population means are of interest,  each Monte Carlo iteration requires summations involving $N$ expensive memory accesses.%\todo{%
%	Do we really want to say $\complexity(N)$ is ``expensive.''
%	Most models / algorithms we present in the article is well beyond $\complexity{N}$.
%}
Another example of a computationally intensive model is Gaussian process regression  \citep{williams1996gaussian,williams2006gaussian};
it is a popular non-parametric approach but the exact method for fitting the model and predicting future values requires matrix inversions that scale $\complexity(N^3)$.
As the rest of the calculations require relatively negligible computational effort, we say that matrix inversions represent the \emph{computational bottleneck} for Gaussian process regression.

\newcommand{\dd}{\mbox{d}}
\newcommand{\x}{\mathbf{x}}
\newcommand{\ttheta}{\boldsymbol\theta}

To speed up a computationally intensive method, one only needs to speed up the method's computational bottleneck.
We are interested in performing Bayesian inference \citep{gelman2013bayesian}  based on a large vector of observations $\x=(x_1, \dots, x_N)$.  We specify our model for the data with a likelihood function $\pi(\x|\ttheta)=\prod_{n=1}^N \pi(x_n|\ttheta)$ and use a prior distribution with density function $\pi(\ttheta)$ to characterize our belief about the value of the $P$-dimensional parameter vector $\ttheta$ \emph{a priori}.  The target of Bayesian inference is the posterior distribution of $\ttheta$ conditioned on $\x$
\begin{align}\label{eq:bayes}
\pi(\ttheta|\x) = \pi(\x|\ttheta) \pi(\ttheta) \;/\, \int \pi(\x|\ttheta) \pi(\ttheta)\, \dd \ttheta \, .
\end{align}
The denominator's multi-dimensional integral quickly becomes impractical as $P$ grows large, so we choose to use the Metropolis-Hastings (M-H) algorithm to generate a Markov chain with stationary distribution $\pi(\ttheta|\x)$ \citep{metropolis1953equation,hastings1970monte,tierney1994markov}.  We begin at an arbitrary position $\ttheta^{(0)}$ and, for each iteration $s = 0,\dots,S$, randomly generate the proposal state $\ttheta^*$ from the transition distribution with density $q(\ttheta^*|\ttheta^{(s)})$. We then accept proposal state $\ttheta^*$ with probability
\begin{align}\label{eq:accrej}
a = \min \left(1, \frac{ \pi(\ttheta^*|\x)
	q(\ttheta^{(s)}|\ttheta^*)}
{\pi(\ttheta^{(s)}|\x)
	q(\ttheta^*|\ttheta^{(s)})}  \right)\, .
\end{align}
The ratio on the right now longer depends on the denominator in Formula \eqref{eq:bayes}, but one must still compute the likelihood and its $N$ terms $\pi(x_n|\ttheta^*)$.

It is for this reason that likelihood evaluations are often the computational bottleneck for Bayesian inference. In the best case, these evaluations are $\complexity(N)$, but there are many situations in which they scale $\complexity(N^2)$ \citep{holbrook2020massive,holbrook2020scalable} or worse.
Indeed, when $P$ is large, it is often advantageous to use more advanced MCMC algorithms that use the gradient of the log-posterior to generate better proposals.  In this situation, the log-likelihood gradient may also become a computational bottleneck \citep{holbrook2020massive}.

\subsection{Big \emph{P}}\label{sec:bigP}

\newcommand{\bbeta}{\boldsymbol\beta}
\newcommand{\X}{\mathbf{X}}
\newcommand{\y}{\mathbf{y}}
\newcommand{\p}{\mathbf{p}}
\newcommand{\M}{\mathbf{M}}

%A canonical model for big $P$ problems is regularized regression, in which the effective number of parameter is controlled by incorporating a penalty term.
%Ridge regression \citep{seber2012linear} is arguably the simplest of such models, but computing becomes expensive with large $P$ even in this classical setting.
One of the simplest models for big $P$ problems is ridge regression \citep{seber2012linear}, but computing can become expensive even in this classical setting.
Ridge regression estimates the coefficient $\ttheta$ by minimizing the distance between observed and predicted values $\y$ and $\X \ttheta$ along with a weighted square norm of $\ttheta$:
\begin{align*}
\hat{\ttheta} 
	= \textrm{argmin} \left\{
		\| \y - \X \ttheta\|^2 + \| \Precision^{1/2} \ttheta \|^2
		\right\}
	= \left(\X^\transposeSymbol \X + \Precision \right)^{-1}\X^\transposeSymbol \y.
\end{align*}
For illustrative purposes, we consider the following direct method for computing $\hat{\ttheta}$.\footnote{
	A more numerically stable approach %based on pivoted QR decomposition 
	has the same complexity \citep{trefethen1997numerical_linalg}.
}
We can first multiply the $N\times P$ design matrix $\X$ by its transpose at the cost of $\complexity(N^2P)$ and subsequently invert the $P \times P$ matrix $(\X^\transposeSymbol \X + \Precision)$ at the cost of $\complexity(P^3)$.
The total $\complexity(N^2P + P^3)$ complexity shows that (1) a large number of parameters is often sufficient for making even the simplest of tasks infeasible and (2) a moderate number of parameters can render a task impractical when there are a large number of observations.  These two insights extend to more complicated models: the same complexity analysis holds for the fitting of generalized linear models (GLMs) as described in \cite{mccullagh1989generalized}.

In the context of Bayesian inference, the length $P$ of the vector $\ttheta$ dictates the dimension of the MCMC state space.  For the M-H algorithm (Section \ref{sec:bigN}) with $P$-dimensional Gaussian target and proposal, \cite{gelman1996efficient} show that the proposal distribution's covariance should be scaled by a factor inversely proportional to $P$.  Hence, as the dimension of the state space grows, it behooves one to propose states $\ttheta^*$ that are closer to the current state of the Markov chain, and one must greatly increase the number $S$ of MCMC iterations.  At the same time, an increasing $P$ often slows down rate-limiting  likelihood calculations (Section \ref{sec:bigN}). Taken together, one must generate many more, much slower MCMC iterations.  The wide-applicability of latent variable models \citep{van2001art} (Sections \ref{sec:sparse} and \ref{sec:phylogen}) for which each observation has its own parameter set (e.g., $P \propto N$) means M-H simply does not work for a huge class of models popular with practitioners.

For these reasons, Hamiltonian Monte Carlo (HMC) \citep{neal2011mcmc} has become a popular algorithm for fitting Bayesian models with large numbers of parameters.  Like M-H, HMC uses an accept-step (Equation \eqref{eq:accrej}). Unlike M-H, HMC takes advantage of additional information about the target distribution in the form of the log-posterior gradient. HMC works by doubling the state space dimension with an auxiliary Gaussian `momentum' variable  $\p\sim \normal_P(\mathbf{0},\M)$ independent to the `position' variable $\ttheta$. The constructed Hamiltonian system has energy function given by the negative logarithm of the joint distribution
\begin{align*}
H(\ttheta,\p) \propto - \log \left( \pi(\ttheta|\X)  \times \exp (-\p^T\M^{-1}\p/2)  \right)  \propto - \log \pi(\ttheta|\X) + \p^T\M^{-1}\p/2 \, ,
\end{align*}
and  we produce proposals by simulating the system according to Hamilton's equations
\begin{align*}
\dot{\ttheta} &= \frac{\partial}{\partial \p}H(\ttheta,\p) =  M^{-1} \p /2\\ \nonumber
\dot{\p} &= - \frac{\partial}{\partial \ttheta}H(\ttheta,\p) = \nabla \log \pi(\ttheta|\X) \, .
\end{align*}
Thus, the momentum of the system moves in the direction of steepest ascent for the log-posterior forming an analogy with first-order optimization. The cost is repeated gradient evaluations that may comprise a new computational bottleneck, but the result is effective MCMC for tens of thousands of parameters \citep{holbrook2017bayesian,holbrook2020massive}. The success of HMC has inspired research into other methods leveraging gradient information to generate better MCMC proposals when $P$ is large \citep{bouchard2018bouncy}.

\subsection{Big \emph{M}}\label{sec:bigM}

\newcommand{\Ggamma}{\boldsymbol\Gamma}

Global optimization, or the problem of finding the minimum of  a function with arbitrarily many local minima, is NP-complete in general \citep{murty1985some}, meaning---in layman's terms---it is impossibly hard.  In the absence of a tractable theory, by which one might prove one's global optimization procedure works, brute-force grid and random searches and heuristic methods such as particle swarm optimization \citep{kennedy1995particle} and genetic algorithms \citep{davis1991handbook} have been popular. Due to the overwhelming difficulty of global optimization, a large portion of the optimization literature has focused on the particularly well-behaved class of \emph{convex} functions \citep{hunter2004tutorial,boyd2004convex}, which do not admit multiple local minima.  Since Fisher introduced his `maximum likelihood' in 1922 \citep{fisher1922mathematical}, statisticians have thought in terms of maximization, but convexity theory still applies by a trivial negation of the objective function.  Nonetheless, most statisticians safely ignored \emph{concavity} during the 20th century: exponential family log-likelihoods are log-concave, so Newton-Raphson and Fisher scoring are guaranteed optimality in the context of GLMs \citep{boyd2004convex,mccullagh1989generalized}.

Nearing the end of the 20th century, multimodality and non-convexity became more important for statisticians considering high-dimensional regression, i.e., regression with many covariates (big $P$).  Here, for purposes of interpretability and variance reduction, one would like to induce \emph{sparsity} on the weights vector $\hat{\ttheta}$ by performing best subset selection \citep{beale1967discarding,hocking1967selection}:
\begin{align}\label{eq:bestsubset}
\hat{\ttheta} = \argmin_{\ttheta \in \mathbb{R}^P}  || \y - \X \ttheta||^2_2  \quad \mbox{subject to} \quad ||\ttheta||_0 \leq k \, ,
\end{align}
where $0<k\ll P$, and $||\cdot||_0$ denotes the $\ell_0$-norm, i.e., the number of non-zero elements.  Because best subset selection requires an immensely difficult non-convex optimization, \cite{tibshirani1996regression} famously replaces the $\ell_0$-norm with the $\ell_1$-norm, thereby providing sparsity while nonetheless maintaining convexity.

Historically, Bayesians have payed much less attention to convexity than have optimization researchers. This is most likely because the basic theory \citep{tierney1994markov} of  MCMC does not require such restrictions: even if a target distribution has one-million modes, the well constructed Markov chain explores them all in the limit.
Despite these theoretical guarantees, a small literature has developed to tackle multi-modal Bayesian inference \citep{geyer1991markov,tjelmeland2001mode,lan2014wormhole,nishimura2016geometrically} because multi-modal target distributions \emph{do} present a challenge in practice.  In analogy with Equation \eqref{eq:bestsubset}, Bayesians seek to induce sparsity by specifiying priors such as the spike-and-slab \citep{mitchell1988bayesian,madigan1994model,george1997approaches}, e.g.,
\begin{align*}
\y \sim \normal_N\left(\X \Ggamma \ttheta ,\sigma^2 \mathbf{I}_N \right) \quad \mbox{for} \quad [\Ggamma]_{\paramIndex \paramIndex'} =  \begin{cases}
\gamma_\paramIndex \sim \bernoulli(\pi) & \paramIndex = \paramIndex' \\
0 & \paramIndex \neq \paramIndex'
\end{cases} \quad \mbox{and} \quad \pi \in (0,1) \, .
\end{align*}
As with the best subset selection objective function, the spike-and-slab target distribution becomes heavily multimodal as $P$ grows and the support of $\Ggamma$'s discrete distribution grows to $2^P$ potential configurations.

In the following section, we present an alternative Bayesian sparse regression approach that mitigates the combinatorial problem along with a state-of-the-art computational technique that scales well both in $N$ and $P$.

\section{Model-specific advances}\label{sec:modSpecific}

These challenges will remain throughout the 21st century, but it is possible to make significant advances for specific statistical tasks or classes of models.
Section \ref{sec:sparse} considers Bayesian sparse regression based on continuous shrinkage priors, designed to alleviate the heavy multimodaility (big $M$) of the more traditional spike-and-slab approach.
This model presents a major computational challenge as $N$ and $P$ grow, but a recent computational advance makes the posterior inference feasible for many modern large-scale applications.

And because of the rise of data science, there are increasing opportunities for computational statistics to grow by enabling and extending statistical inference for scientific applications previously outside of mainstream statistics.  Here, the science may dictate the  development of structured models with complexity possibly growing in $N$ and $P$.  Section \ref{sec:phylogen} presents a method for fast phylogenetic inference, where the primary structure of interest is a `family tree' describing a biological evolutionary history.

\subsection{Bayesian sparse regression in the age of big \emph{N} and big \emph{P}}
\label{sec:sparse}

With the goal of identifying a small subset of relevant features among a large number of potential candidates, sparse regression techniques have long featured in a range of statistical and data science applications \citep{hastie2015learning-with-sparsity}.
Traditionally, such techniques were commonly applied in the ``$N \ll P$'' setting and correspondingly computational algorithms focused on this situation \citep{friedman2010coord-descent}, especially within the Bayesian literature \citep{bhattacharya2016fast_sampling}.

Due to a growing number of initiatives for large-scale data collections and new types of scientific inquiries made possible by emerging technologies, however, increasingly common are datasets that are ``big $N$'' and ``big $P$'' at the same time.
For example, modern observational studies using healthcare databases routinely involve $N \approx 10^5 \sim 10^6$ patients and $P \approx 10^4 \sim 10^5$ clinical covariates \citep{suchard2019legend}.
%For example, in modern observational studies using healthcare databases, the number of patients $N$ and of clinical covariates $P$ are often in the order of $10^5 \sim 10^6$ and $10^4 \sim 10^5$ \citep{suchard2019legend}.
The UK Biobank provides brain imaging data on $N = 100{,}000$ patients, with $P = 100 \sim 200{,}000$ depending on the scientific question of interests  \citep{passos2019personalized_psychiatry}.
Single cell RNA sequencing can generates datasets with $N$ (the number of cells) in millions and $P$ (the number of genes) in tens of thousands, with the trend indicating further growths in data size to come \citep{svensson2019curated}

\subsubsection*{Continuous shrinkage: alleviating big \emph{M}}

Bayesian sparse regression, despite its desirable theoretical properties and flexibility to serve as a building block for richer statistical models,
%advantages over more computationally convenient alternatives such as Lasso \citep{bhadra2017lasso-meets-horseshoe},
has always been relatively computationally intensive even before the advent of ``big $N$ and big $P$'' data \citep{george1997approaches, nott2005adaptive_sampling_for_variable_selection, ghosh2011design_orthogonalization}.
A major source of its computational burden is severe posterior multi-modality (big $M$) induced by the discrete binary nature of spike-and-slab priors (Section \ref{sec:bigM}).
The class of \textit{global-local} continuous shrinkage priors is a more recent alternative to shrink $\theta_p$s in a more continuous manner, thereby alleviating (if not eliminating) the multi-modality issue \citep{carvalho2010horseshoe, polson2010global_local}.
This class of prior is represented as a scale-mixture of Gaussians:
\begin{equation*}
\regcoef_\paramIndex \given \lscale_\paramIndex, \gscale
\sim \normal_N(0, \gscale^2 \lscale_\paramIndex^2), \
\lscale_\paramIndex \sim \localPrior, \
\gscale \sim \globalPrior.
\end{equation*}
The idea is that the \textit{global scale} parameter $\gscale \ll 1$ would shrink most $\regcoef_\paramIndex$'s towards zero while  \textit{local scale} $\lscale_\paramIndex$'s, with its heavy-tailed prior $\localPrior$, allow a small number of $\gscale \lscale_\paramIndex$ and hence $\theta_\paramIndex$'s to be estimated away from zero.
While motivated by two different conceptual frameworks, the spike-and-slab can be viewed as a subset of global-local priors in which $\localPrior$ is chosen as a mixture of delta masses placed at $\lscale_\paramIndex = 0$ and $\lscale_\paramIndex = \sigma / \gscale$.
Continuous shrinkage mitigates the multi-modality of spike-and-slab by smoothly bridging small and large values of $\lscale_\paramIndex$.

On the other hand, use of continuous shrinkage priors does not address the increasing computational burden from growing $N$ and $P$ in modern applications.
Sparse regression posteriors under global-local priors are amenable to an effective Gibbs sampler, a popular class of MCMC we describe further in Section~\ref{sec:genalg}.
%When the observed data assumes a distribution $\y \given \X, \bregcoef, \bOmega \sim \normal_N(\X \bregcoef, \bOmega^{-1})$ conditional on the additional parameter $\bOmega$ --- which is the case in linear and logistic regression\footnote{%
%	The logistic likelihood can be expressed as a Gaussian distribution conditional on an auxiliary P{\'o}lya-Gamma parameter \citep{polson2013polya_gamma}; see \cite{nishimura2018cg_accelerated_gibbs}, for example.
%}
%---
Under the linear and logistic models, the computational bottleneck of this Gibbs sampler stems from the need for repeated updates of $\bregcoef$ from its conditional distribution
\begin{equation}
\label{eq:regcoef_posterior_conditional}
\bregcoef \given \gscale, \blscale, \bm{\Omega}, \y, \X
	\sim \normal_P(\Precision^{-1}\X^\transposeSymbol \bOmega \y, \Precision^{-1}))
	\ \text{ for } \
	\Precision = \X^\transposeSymbol \bOmega \X + \tau^{-2} \bLscale^{-2},
\end{equation}
where $\bOmega$ is an additional parameter of diagonal matrix and $\bLscale = \text{diag}(\blscale)$.\footnote{%
 % is a diagonal matrix with entries $\Lscale_{\paramIndex \paramIndex} = \lscale_\paramIndex$.\footnote{%
	The matrix parameter $\bOmega$ coincides with $\bOmega = \sigma^{-2} \mathbf{I}_N$ for linear regression and $\bOmega = \textrm{diag}(\bomega)$ for auxiliary P{\'o}lya-Gamma parameter $\bomega$  for logistic regression \citep{polson2013polya_gamma, nishimura2018cg_accelerated_gibbs}.
}
Sampling from this high-dimensional Gaussian distribution requires $\complexity(NP^2 + P^3)$ operations with the standard approach \citep{rue2005gmrf}: $\complexity(NP^2)$ for computing the term $\X^\transposeSymbol \bOmega \X$ and $\complexity(P^3)$ for Cholesky factorization of $\Precision$.
While an alternative approach by \cite{bhattacharya2016fast_sampling} provides the complexity of $\complexity(N^2 P + N^3)$, the computational cost remains problematic in the ``big $N$ and big $P$'' regime at $\complexity(\min\{N^2 P, N P^2\})$ after choosing the faster of the two.

\subsubsection*{Conjugate gradient sampler for structured high-dimensional Gaussians}
The \textit{conjugate gradient (CG) sampler} of \cite{nishimura2018cg_accelerated_gibbs} combined with their \textit{prior-preconditioning} technique overcomes this seemingly inevitable $\complexity(\min\{N^2 P, N P^2\})$ growth of the computational cost.
Their algorithm is based on a novel application of the CG method \citep{hestenes1952cg, lanczos1952iterative}, which belongs to a family of \textit{iterative methods} in numerical linear algebra.
Despite its first appearance in 1952, CG received little attention for the next few decades, only making its way into major software packages such as \textsc{matlab} in 1990s \citep{vorst2003iterative}. % Section 8.3
With its ability to solve a large and structured linear system $\Precision \bregcoef = \bm{b}$ via a small number of matrix-vector multiplications $\bm{v} \to \Precision \bm{v}$ without ever explicitly inverting $\Precision$, however, CG has since  emerged as an essential and prototypical algorithm for modern scientific computing \citep{cipra2000top10algo, dongarra2016hpc_connjugate_gradient}.

Despite its earlier rise to prominence in other fields, CG has not found practical applications in Bayesian computation until rather recently \citep{nishimura2018cg_accelerated_gibbs, zhang2019practical-scalable-computing}.
We can offer at least two explanations for this.
First, being an algorithm for solving a deterministic linear system, it is not obvious how CG would be relevant to Monte Carlo simulation, such as sampling from $\normal_P(\bm{\mu}, \Precision^{-1})$; ostensively, such a task requires computing a ``square root'' $\CholeskyFactor$ of the precision matrix so that $\Var(\CholeskyFactor^{-1} \bm{z}) = \CholeskyFactor^{-1} \CholeskyFactor^{-\transposeSymbol} = \Precision^{-1}$ for $\bm{z} \sim \normal_P(\bm{0}, \bm{I}_P)$.
Secondly, unlike direct linear algebra methods, iterative methods such as CG  have a variable computational cost that depends critically on the user's choice of a preconditioner and thus \emph{cannot} be used as a ``black-box'' algorithm.\footnote{%
	See \citealt{nishimura2018cg_accelerated_gibbs} and references therein for the role and design of a preconditioner.
}
In particular, this novel application of CG to Bayesian computation is a reminder that other powerful ideas in other computationally intensive fields may remain untapped by the statistical computing community;
knowledge transfers will likely be facilitated by having more researchers working at intersections of different fields.

\cite{nishimura2018cg_accelerated_gibbs} turns CG into a viable algorithm for Bayesian sparse regression problems by realizing that 1) we can obtain a Gaussian vector $\bm{b} \sim \normal_P\big( \X^\transposeSymbol \bOmega \y, \Precision \big)$ by first generating $\bm{z} \sim \normal_P(\bm{0}, \bm{I}_P)$ and $\bm{\zeta} \sim \normal_N(\bm{0}, \bm{I}_N)$ and then setting $\bm{b} = \X^\transposeSymbol \bOmega \y + \X^\transposeSymbol \bm{\Omega}^{1/2} \bm{\zeta} + \tau^{-1} \bLscale^{-1} \bm{z}$, and 2) subsequently solving $\Precision \bregcoef = \bm{b}$ yields a sample $\bregcoef$ from the distribution \eqref{eq:regcoef_posterior_conditional}.
The authors then observe that the mechanism through which a shrinkage prior induces sparsity of $\regcoef_\paramIndex$'s also induces a tight clustering of eigenvalues in the prior-preconditioned matrix $\gscale^{2} \bLscale \Precision \bLscale$.
This fact makes it possible for prior-preconditioned CG to solve the system $\Precision \bregcoef = \bm{b}$ in $K$ matrix-vector operations of form $\bm{v} \to \Precision \bm{v}$, where $K$ roughly represents the number of ``significant'' $\regcoef_\paramIndex$'s that are distinguishable from zeros under the posterior.
%$\Precision = \X^\transposeSymbol \bOmega \X + \tau^{-2} \bLscale^{-2}$,
For $\Precision$ having a structure as in \eqref{eq:regcoef_posterior_conditional},
$\Precision \bm{v}$ can be computed via matrix-vector multiplications of form $\bm{v} \to \X \bm{v}$ and $\bm{w} \to \X^\transposeSymbol \bm{w}$, so each $\bm{v} \to \Precision \bm{v}$ operation requires a fraction of the computational cost of directly computing $\Precision$ and then factorizing it.
%This clustering of eigenvalues causes a rapid convergence of when applied to

Prior-preconditioned CG demonstrates an order of magnitude speed-up in posterior computation when applied to a comparative effectiveness study of atrial fibrillation treatment involving $N = 72{},489$ patients and $P = 22{,}175$ covariates \citep{nishimura2018cg_accelerated_gibbs}.
Though unexplored in their work, the algorithm's heavy use of matrix-vector multiplications provides avenues for further acceleration.
Technically, the algorithm's complexity may be characterized as $\complexity(NPK)$, for the $K$ matrix-vector multiplications by $\X$ and $\X^\transposeSymbol$, but the theoretical complexity is only a part of the story.
Matrix-vector multiplications are amenable to a variety of hardware optimizations, which in practice can make orders of magnitude difference in speed (Section~\ref{sec:hardware}).
In fact, given how arduous manually optimizing computational bottlenecks can be, designing algorithms so as to take advantage of common routines (as those in Level 3 \textsc{blas}) and their ready-optimized implementations has been recognized as an effective principle in algorithm design \citep{golub2012matrix_computation}.
%Incidentally, given the prevalence of sparse binary design matrices in modern large-scale applications \citep{friedman2010coord-descent, nishimura2018cg_accelerated_gibbs}, the authors would like to advocate

\newcommand{\br}[1]{{r}_{#1}}
\newcommand{\treeMean}{\mu}
\newcommand{\randomEffect}[1]{\epsilon_{#1}}
\newcommand{\REvariance}{\psi}

\subsection{Phylogenetic reconstruction}\label{sec:phylogen}

\begin{figure}[ht!]
	\begin{center}
		\includegraphics[width=0.9\linewidth]{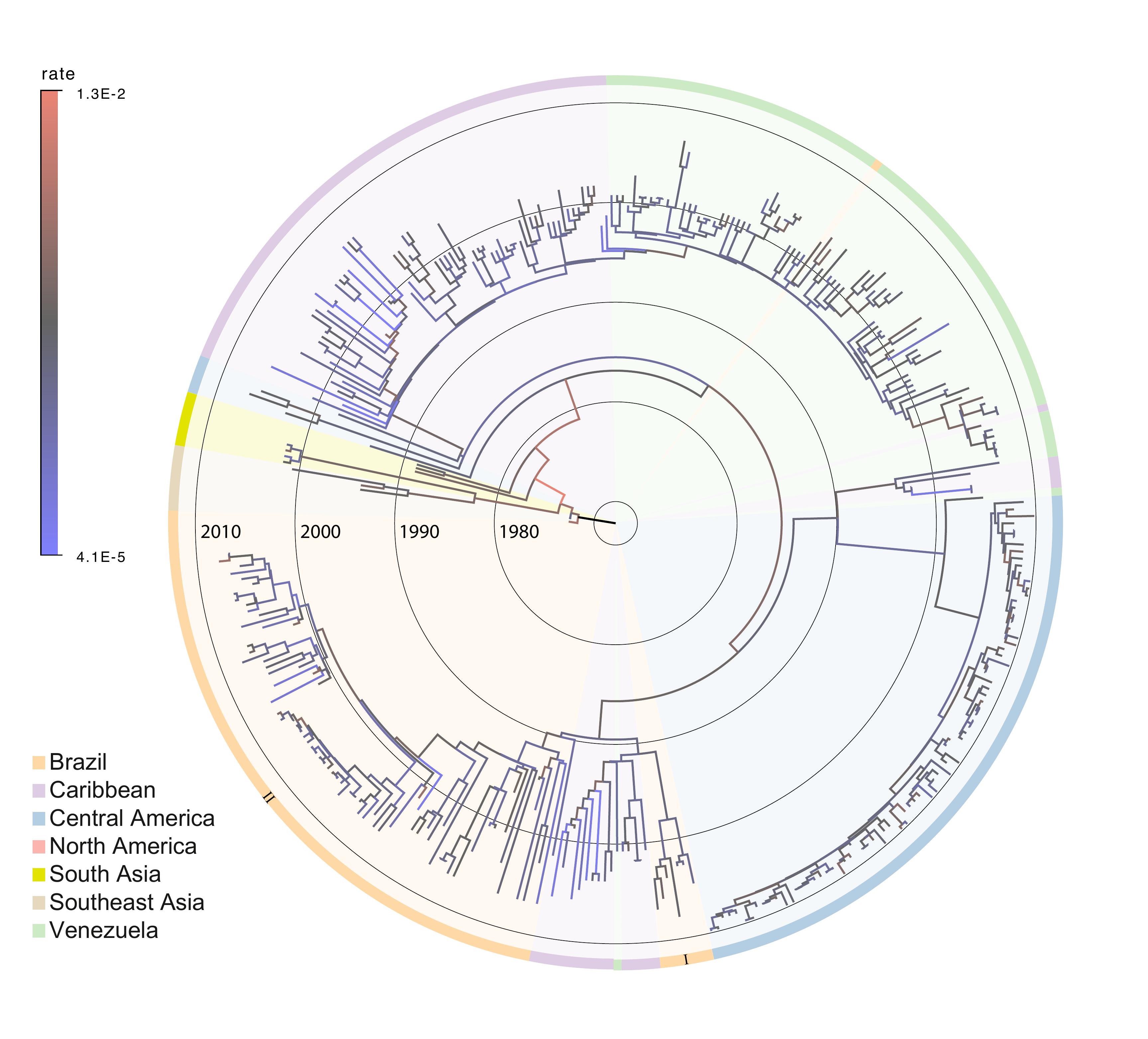}
	\end{center}
	\caption{
		A non-traditional and critically important application in computational statistics is the reconstruction of evolutionary histories in the form of phylogenetic trees.
		Here is a maximum clade credible tree of the Dengue virus example.
		The dataset consists of $352$ sequences of the serotype $3$ of the Dengue virus.
		Branches are color-coded by the posterior means of the branch-specific evolutionary rates according to the color bar on the top left.
		The concentric circles indicate the time scale with the year numbers.
		%PL: something funky seems to be going on ar the root?
		%XJ: yes, MAS asked the same question.  There are actually two tip nodes...  They look less funky in the rectangular tree and appear in the original publication too, but with slightly different sub-tee topology.
		%PL: OK, i have committed a version, LargeDengue_tree_logRate2.pdf, with a slight edit at the root so that it is more clear that there are two tips (and not a box-like thing) -- have a look
		The outer ring indicates the geographic locations of the samples by the color code on the bottom left.
		`\textbf{I}' and `\textbf{II}' indicate the two Brazilian lineages as in the original study.
	}
	\label{fig:DengueTree}
\end{figure}

Whereas big $N$ and big $P$ regression adapts a classical statistical task to contemporary needs, the 21st century is witnessing the application of computational statistics to the entirety of applied science.
One such example is the tracking and reconstruction of deadly global viral pandemics.
Molecular phylogenetics has become an essential analytical tool for understanding the complex patterns in which rapidly evolving pathogens propagate throughout and between countries, owing to the complex travel and transportation patterns evinced by modern economies \citep{Pybus2015}, along with other factors such as increased global population and urbanisation \citep{bloom2017}.
The advance in sequencing technology is generating pathogen genomic data at an ever-increasing pace, with a trend to real-time that requires the development of computational statistical methods that are able to process the sequences in a timely manner and produce interpretable rsults to inform national/global public health organizations.

The previous three Core Challenges are usually interwind such that the increase in the sample size (big N) and the number of traits (big P) for each sample usually happen simultaneously and lead to increased heterogeneity that requires more complex models (big M).
For example, recent studies in viral evolution have seen 
a continuing increase in the sample size that the West Nile virus, Dengue, HIV, Ebola virus studies involve $104$, $352$, $465$, $1610$ sequences \citep{Pybus2012, nunes2014air, bletsa2019divergence, dudas2017virus} and the GISAID database has collected $92,000$ COVID-19 genomic sequences by the end of August 2020 \citep{elbe2017data}.
% Not surprisingly, the improvement to solving the challenges often come with a combination of algorithmic and computational advances.

\newcommand{\iterPerUpdate}{k}
\newcommand{\position}{\boldsymbol{\theta}}
To accomodate the increasing size and heterogeneity in the data and be able to apply the aforementioned efficient gradient-based algorithms, \cite{ji2020gradients} propose a linear-time algorithm for calculating an $O(N)$-dimensional gradient on a tree w.r.t.~the sequence evolution.
The linear-time gradient algorithm calculates each branch-specific derivative through a pre-order traversal that complements the post-order traversal from the likelihood calculation of the observed the sequence data at the tip of the phylogeny by marginalizing over all possible hidden states on the internal nodes.
The pre- and post-order traversals complete the Baum's forward-backward algorithm in a phylogenetic framework \citep{Baum1972}.
The authors then apply the gradient algorithm with HMC (Section \ref{sec:bigP}) samplers to learn the branch-specific viral evolutionary rates.

Thanks to these advanced computational methods, one can employ more flexible models that lend themselves to more realistic reconstructions and uncertainty quantification.
Following a random-effects relaxed clock model, they model the evolutionary rate $\br{p}$ of branch $p$ on a phylogeny as the product of a global tree-wise mean parameter $\treeMean$ and a branch-specific random effect $\randomEffect{p}$.
They model the random-effect $\randomEffect{p}$'s as independent and identically distributed from a lognormal distribution such that $\randomEffect{p}$ has mean $1$ and variance $\REvariance^2$ under a hierarchical model where $\REvariance$ is the scale parameter.
To accomodate the difference in scales of the variability in the parameter space for the HMC sampler, the authors adopt preconditioning with adaptive mass matrix informed by the diagonal entries of the Hessian matrix.
More precisely, the non-zero diagonal elements of the mass matrix truncate the values from the first $s$ HMC iterations of
$H_{pp}^{(s)} =
	\frac{1}{\lfloor s / \iterPerUpdate \rfloor} \sum_{s \, : \, s / \iterPerUpdate \, \in \, \mathbb{Z}^+}
	\left. \left[ - \frac{\partial^2  }{\partial^2 \theta_p} \log \pi(\position) \right|_{\position=\position^{(s)}} \right]
	\approx \mathbb{E}_{\pi(\theta)} \left[
		- \frac{\partial^2}{\partial^2 \theta_i} \log \pi(\position)
	\right]$
so that the matrix remains positive-definite and numerically stable.
They estimate the tree-wise (fixed-effect) mean rate $\treeMean$ with posterior mean $4.75$ ($95\%$ Bayesian credible interval : $4.05, 5.33$) $\times 10^{-4}$ substitutions per site per year with rate variability characterized by scale parameter with posterior mean $\REvariance=1.26[1.06, 1.45]$ for serotype 3 of Dengue virus with a sample size of $352$ \citep{nunes2014air}.
Figure~\ref{fig:DengueTree} illustrates the estimated maximum clade credible evolutionary tree of the Dengue virus dataset.

The authors report relative speedup in terms of the effective sample size per second (ESS/s) of the HMC samplers compared to a univariate transition kernal.
The `vanilla' HMC sampler with an identity mass matrix gains $2.2\times$ speedup for the minimum ESS/s and $2.5\times$ speedup for the median ESS/s whereas the `preconditioned' HMC sampler gains $16.4\times$ and $7.4\times$ speedups respectively.  Critically, the authors make these performance gains available to scientists everywhere through the popular, open-source software package for viral phylogenetic inference \emph{Bayesian evolutionary analysis by sampling trees} (\textsc{BEAST}) \citep{suchard2018bayesian}.  In Section \ref{sec:genalg}, we discuss how software like \textsc{BEAST} address Core Challenge 4, the creation of fast, flexible and friendly statistical algo-ware.

\section{Core challenges 4 and 5}

Section \ref{sec:modSpecific} provides examples of how computational statisticians might address Core Challenges 1-3 (big $N$, big $P$ and big $M$) for individual models.
Such advances in computational methods must be accompanied by easy-to-use software to make them accessible to end-users. As \cite{gentle2012computational} puts it, ``While referees and editors of scholarly journals determine what statistical theory and methods are published, the developers of the major statistical software packages determine what statistical methods are used.''
We would like statistical software to be widely applicable yet computationally efficient at the same time.
Trade-offs invariably arise between these two desiderata, but one should nonetheless strive to design algorithms that are general enough to solve an important class of problems and as efficiently as possible in doing so.

Section \ref{sec:genalg} presents Core Challenge 4, achieving `algo-ware' (a neologism suggesting an equal emphasis on the statistical algorithm and its implementation) that is sufficiently efficient, broad and user-friendly to empower everyday statisticians and data scientists.
Core Challenge 5 (Section \ref{sec:hardware}) explores the mapping of these algorithms to computational hardware for optimal performance.
Hardware-optimized implementations often exploit model-specific structures, but good, general-purpose software should also optimize common routines.

\subsection{Fast, flexible and friendly statistical algo-ware}\label{sec:genalg}

\newcommand{\Ssigma}{\boldsymbol{\Sigma}}

To accommodate the greatest range of models while remaining simple enough to encourage easy implementation, inference methods should rely solely on the quantities that can be computed algorithmically for any given model.
The log-likelihood (or log-density in the Bayesian setting) is one such quantity, and one can employ the computational graph framework \citep{lunn2009bugs, bergstra2010theano} to evaluate conditional log-likelihoods for any subset of model parameters, as well as their gradients via back propagation \citep{rumelhart1986learning}.  Beyond being efficient in terms of the first three Core Challenges, an algorithm should demonstrate robust performance on a reasonably wide range of problems without extensive tuning if it is to lend itself to successful software deployment.

HMC (Section \ref{sec:bigP}) is a prominent example of a general-purpose algorithm for Bayesian inference, only requiring the log-density and its gradient.
The generic nature of HMC has opened up possibilities for complex Bayesian modeling as early as \cite{neal1996bayesian_neuralnet}, but its performance is highly sensitive to model parametrization and its three tuning parameters, commonly referred to as trajectory length, stepsize and mass matrix \citep{neal2011mcmc}.  Tuning issues constitute a major obstacle to the wider adoption of the algorithm, as evidenced by the development history of the popular HMC-based probabilistic programming software Stan \citep{gelman2014petascale_grant_report}, which employs the \emph{No-U-Turn} (NUTS) sampler of \cite{hoffman2014nuts} to make HMC user-friendly by obviating the need to tune its trajectory length.
Bayesian software packages such as Stan empirically adapt the remaining stepsize and mass matrix \citep{stan2018}; this approach helps make the use of HMC automatic though is not without issues \citep{livingstone2019robustness} and comes at the cost of significant computational overhead.

%Despite all the caveat, the emergence of general-purpose Bayesian inference software powered by HMC is
Although HMC is a powerful algorithm that has played a critical role in the emergence of general-purpose Bayesian inference software, the challenges involved in its practical deployment also demonstrate how an algorithm --- no matter how versatile and efficient at its best --- is not necessarily useful unless it can be made easy for practitioners to use.
It is also unlikely that one algorithm works well in all situations.
In fact, there are many distributions on which HMC performs poorly \citep{stan2018, mangoubi2018hmc_mixing_on_multimodal, livingstone2019kinetic_energy_choice}.
Additionally, HMC is incapable of handling discrete distributions in a fully general manner despite the progresses made in extending HMC to such situations \citep{dinh2017probabilistic-hmc, nishimura2020discontinuous}.
%For these reasons, specialized algorithms that solve and are optimized for important classes of models, as those presented in Section~\ref{sec:modSpecific}, remain highly relevant for statistical software.

But broader applicability comes with its own challenges.  Among sampling-based approaches to Bayesian inference, the Gibbs sampler \citep{geman1984stochastic, gelfand1990sampling} is, arguably, the most versatile of MCMC methods.
The algorithm simplifies the task of dealing with a complex multi-dimensional posterior distribution by factorizing the posterior into simpler conditional distributions for blocks of parameters and iteratively updating parameters from their conditionals.
Unfortunately, the efficiency of an individual Gibbs sampler depends on its specific factorization and the degree of dependence between its blocks of parameters.
Without a careful design or in the absence of effective factorization, therefore, Gibbs samplers' performance may lag behind alternatives such as HMC \citep{monnahan2017faster}.

On the other hand, Gibbs samplers often require little tuning and can take advantage of highly optimized algorithms for each conditional update, as done in the examples of Section~\ref{sec:modSpecific}. A clear advantage of the Gibbs sampler is that it tends to make software implementation quite modular; for example, each conditional update can be replaced with the latest state-of-the-art samplers as they appear \citep{zhang2019large}, and adding a new feature may amount to no more than adding a single conditional update \citep{suchard2018bayesian}.
In this way, an algorithm may not work in a completely model-agnostic manner but with a broad enough scope can serve as a valuable recipe or meta-algorithm for building model-specific algorithms and software.
The same is true for optimization methods. Even though its ``E''-step requires a derivation (by hand) for each new model, the EM algorithm \citep{dempster1977maximum} enables maximum likelihood estimation for a wide range of models.
Similarly, variational inference (VI) for approximate Bayes requires manual derivations but provides a general framework to turn posterior computation into an optimization problem \citep{jordan1999variational}.
As meta-algorithms, both EM and VI expand their breadth of use by replacing analytical derivations with Monte Carlo estimators but suffer losses in statistical and computational efficiency \citep{wei1990monte, ranganath2014black}.
% Two other popular optimization techniques win with ease and breadth of use: the alternating directions method of multipliers (ADMM) \citep{boyd2011distributed} is a simple to implement convex optimization algorithm (Section \ref{sec:bigM}) that has the modern benefit of being easy to parallelize across disparate datasets; and stochastic gradient descent with back propagation scales deep neural networks to big data by evaluating a loss function's gradient over only a small batch of data points at a time.
Indeed, such trade-offs will continue to haunt the creation of fast, flexible and friendly statistical algo-ware well into the 21st century.

\subsection{Hardware optimized inference}\label{sec:hardware}

%\begin{itemize}
%	\item hardware: milti-core, simd, gpus, tpus, quantum
%	\item optimized data movement
%	\item Aki's actually efficient algorithms
%\end{itemize}

\newcommand{\bb}{\mathbf{b}}
\newcommand{\MM}{\mathbf{M}}
\newcommand{\uu}{\mathbf{u}}
\newcommand{\q}{\mathbf{q}}

But successful statistical inference software must also interact with computational hardware in an optimal manner.  Growing datasets require the computational statistician to give more and more thought to how the computer implements any statistical algorithm.  To effectively leverage computational resources, the statistician must (1) identify the routine's computational bottleneck (Section \ref{sec:bigN}) and (2) algorithmically map this rate-limiting step to available hardware such as a multi-core or vectorized central processing unit (CPU), a many-core graphics processing unit (GPU) or---in the future---a quantum computer.
Sometimes, the first step is clear theoretically: a naive implementation of the high-dimensional regression example of Section \ref{sec:sparse} requires an order $\complexity(N^2P)$ matrix multiplication followed by an order $\complexity(P^3)$ Cholesky decomposition. Other times, one can use an instruction-level program profiler, such as \textsc{Intel VTune} (Windows, Linux) or \textsc{Instruments} (OSX), to identify a performance bottleneck. Once the bottleneck is identified one must choose between computational resources, or some combination thereof, based on relative strengths and weaknesses as well as natural parallelism of the target task.

Multi-core CPU processing is effective for parallel completion of multiple, mostly independent tasks that do not require inter-communication.  One might generate 2 to, say, 72 independent Markov chains on a desktop computer or shared cluster.  A positive aspect is that the tasks do not have to involve the same instruction sets at all; a negative is \emph{latency}, i.e., that the slowest process dictates overall runtime.  It is possible to further speedup CPU computing with single instruction, multiple data (SIMD) or vector processing.  A small number of vector processing units (VPU) in each CPU core can carry out a single set of instructions on data stored within an extended-length register.  Intel's streaming SIMD extensions (SSE), advance vector extensions (AVX) and AVX-512 allow operations on 128, 256 and 512 bit registers. In the context of 64 bit double precision, theoretical speedups for SSE, AVX and AVX-512 are 2-fold, 4-fold and 8-fold.  For example, if a computational bottleneck exists within a for-loop, one can unroll the loop and perform operations on, say, 4 consecutive loop bodies at once using AVX \citep{holbrook2020massive,holbrook2020scalable}.  Conveniently, languages like \textsc{OpenMP} \citep{dagum1998openmp} make SIMD loop optimization transparent to the user \citep{warne2019acceleration}.  Importantly, SIMD and multi-core optimization play well together, providing multiplicative speedups.

Whereas a CPU may have tens of cores, GPUs accomplish fine-grained parallelization with thousands of cores that apply a single instruction set to distinct data within smaller work-groups of tens or hundreds of cores.  Quick communication and shared cache-memory within each work-group balance full parallelization across groups, and dynamic on-loading and off-loading of the many tasks hides the latency that is so problematic for multi-core computing.  Originally designed for efficiently parallelized matrix math calculations arising from image rendering and transformation, GPUs easily speed up tasks that are tensor multiplication intensive such as deep learning \citep{bergstra2011theano}, but general-purpose GPU applications abound.  \cite{holbrook2020massive} provides a larger review of parallel computing within computational statistics.  The same paper reports a GPU providing 200-fold speedups over single-core processing and 10-fold speedups over 12-core AVX processing for likelihood and gradient calculations while sampling from a Bayesian multi-dimensional scaling posterior using HMC at scale.  \cite{holbrook2020scalable} reports similar speedups for inference based on spatio-temporal Hawkes processes. Neither application involves matrix or tensor manipulations.

A quantum computer acts on complex data vectors of magnitude 1 called qubits with gates that are mathematically equivalent to unitary operators \citep{nielsen2002quantum}.  Assuming that engineers overcome the tremendous difficulties involved in building a practical quantum computer (where practicality entails simultaneous use of many quantum gates with little additional noise), 21st century statisticians might have access to quadratic or even exponential speedups for extremely specific statistical tasks.  We are particularly interested in the following five quantum algorithms: quantum search \citep{grover1996fast}, or finding a single 1 amid a collection of 0s, only requires $\complexity(\sqrt{N})$ queries, delivering a quadratic speedup over classical search; quantum counting \citep{boyer1998tight}, or finding the number of 1s amid a collection of 0s, only requires $\complexity(\sqrt{N/M})$ (where $M$ is the number of 1s) and could be useful for generating p-values within Monte Carlo simulation from a null distribution (Section \ref{sec:bigN}); the quantum minimization algorithm of \citet{durr1996quantum} finds the minimizer of a finite ordered set in time $\complexity(\sqrt{N})$ and features in the quantum parallel MCMC of \citet{holbrook2021quantum}; to obtain the gradient of a function (e.g., the log-likelihood for Fisher scoring or HMC) with a quantum computer, one only needs to evaluate the function once \citep{jordan2005fast} as opposed to $\complexity(P)$ times for numerical differentiation; finally, the HHL algorithm \citep{harrow2009quantum} obtains the scalar value $\q^T\MM \q$ for the $P$-vector $\q$ satisfying $\mathbf{A}\q=\bb$ and $\MM$ an $P\times P$ matrix in time $\complexity\left(\log (P\kappa^2)\right)$, delivering an exponential speedup over classical methods. Technical caviats exist \citep{aaronson2015read}, but HHL may find use within high-dimensional hypothesis testing (big $P$). Under the null hypothesis, one can rewrite the score test statistic
\begin{align*}
\uu^T(\hat{\ttheta}_0)\, \mathcal{I}^{-1}(\hat{\ttheta}_0)\, \uu(\hat{\ttheta}_0) \quad \mbox{as} \quad \uu^T(\hat{\ttheta}_0)\, \mathcal{I}^{-1}(\hat{\ttheta}_0)\, \mathcal{I}(\hat{\ttheta}_0)\,   \mathcal{I}^{-1}(\hat{\ttheta}_0)\, \uu(\hat{\ttheta}_0)
\end{align*}
for $\mathcal{I}(\hat{\ttheta}_0)$ and $\uu(\hat{\ttheta}_0)$ the Fisher information and log-likelihood gradient evaluated at the maximum likelihood solution under the null hypothesis. Letting $\mathbf{A}=\mathcal{I}(\hat{\ttheta}_0)=\M$ and $\bb=\uu(\hat{\ttheta}_0)$, one may write the test statistic as $\q^T \M \q$ and obtain it in time logarithmic in $P$.  When the model design matrix $\X$ is sufficiently sparse---a common enough occurrence in large-scale regression---to render $\mathcal{I}(\hat{\ttheta}_0)$ itself sparse, the last criterion for the application of the HHL algorithm is met.

%Software: \textsc{Threading Building Blocks} (\textsc{TBB}) and \textsc{RcppParallel}, \textsc{OpenCL}, \textsc{Xsimd}, \textsc{RcppXsimd}, \textsc{RcppNT2}

\section{Rise of data science}

Core Challenges 4 and 5---fast, flexible and user-friendly algo-ware and hardware optimized inference---embody an increasing emphasis on application and implementation in the age of data science. Previously undervalued contributions in statistical computing, e.g., hardware utilization, database methodology, computer graphics, statistical software engineering and the human-computer interface \citep{gentle2012computational}, are slowly taking on greater importance within the (rather conservative) discipline of statistics.  There is perhaps no better illustration of this trend than Dr.~Hadley Wickham's winning the prestigious COPSS Presidents' Award for 2019
\begin{quote}
[for] influential work in statistical computing, visualization, graphics, and data analysis; for developing and implementing an impressively comprehensive computational infrastructure for data analysis through R software; for making statistical thinking and computing accessible to large audience; and for enhancing an appreciation for the important role of statistics among data scientists \citep{copss2019}.
\end{quote}
This success is all the more impressive because Presidents' Awardees have historically been contributors to statistical theory and methodology, not Dr.~Wickham's scientific software development for data manipulation \citep{wickham2007reshaping,wickham2011split,wickham2014tidy} and visualization \citep{kahle2013ggmap,wickham2016ggplot2}.

All of this might lead one to ask: \emph{does the success of data science portend the declining significance of computational statistics and its Core Challenges?}  Not at all!  At the most basic level, data science's emphasis on application and implementation underscores the need for computational thinking in statistics.  Moreover, the scientific breadth of data science brings new applications and models to the attention of statisticians, and these models may require or inspire novel algorithmic techniques.  Indeed, we look forward to a golden age of computational statistics, in which statisticians labor within the intersections of mathematics, parallel computing, database methodologies and software engineering with impact on the entirety of the applied sciences.  After all, significant progress toward conquering the Core Challenges of computational statistics requires that we use every tool at our collective disposal.

\section*{Acknowledgments}

AJH is supported by NIH grant K25AI153816.
MAS is supported by NIH grant U19AI135995 and NSF grant DMS1264153.

\bibliographystyle{sysbio}
\DeclareRobustCommand{\disambiguate}[3]{#1}
\bibliography{refs}

\begin{thebibliography}{113}
\providecommand{\natexlab}[1]{#1}
\providecommand{\selectlanguage}[1]{\relax}
\providecommand{\bibAnnoteFile}[1]{%
  \IfFileExists{#1}{\begin{quotation}\noindent\textsc{Key:} #1\\
  \textsc{Annotation:}\ \input{#1}\end{quotation}}{}}
\providecommand{\bibAnnote}[2]{%
  \begin{quotation}\noindent\textsc{Key:} #1\\
  \textsc{Annotation:}\ #2\end{quotation}}

\bibitem[{Aaronson(2015)}]{aaronson2015read}
Aaronson, S. 2015. Read the fine print. Nature Physics 11:291--293.
\bibAnnoteFile{aaronson2015read}

\bibitem[{{American Statistical Association}(2020)}]{ASA}
{American Statistical Association}. 2020. {Statistics Degrees total and by
  gender}. \url{https://ww2.amstat.org/misc/StatTable1987-Current.pdf} online;
  accessed 01 June 2020.
\bibAnnoteFile{ASA}

\bibitem[{Baum(1972)}]{Baum1972}
Baum, L. 1972. An inequality and associated maximization technique in
  statistical estimation of probabilistic functions of a {Markov} process.
  Inequalities 3:1--8.
\bibAnnoteFile{Baum1972}

\bibitem[{Beale et~al.(1967)Beale, Kendall, and Mann}]{beale1967discarding}
Beale, E., M.~Kendall, and D.~Mann. 1967. The discarding of variables in
  multivariate analysis. Biometrika 54:357--366.
\bibAnnoteFile{beale1967discarding}

\bibitem[{Bergstra et~al.(2011)Bergstra, Bastien, Breuleux, Lamblin, Pascanu,
  Delalleau, Desjardins, Warde-Farley, Goodfellow, Bergeron
  et~al.}]{bergstra2011theano}
Bergstra, J., F.~Bastien, O.~Breuleux, P.~Lamblin, R.~Pascanu, O.~Delalleau,
  G.~Desjardins, D.~Warde-Farley, I.~Goodfellow, A.~Bergeron, et~al. 2011.
  Theano: Deep learning on gpus with python. Pages~1--48 \emph{in} NIPS 2011,
  BigLearning Workshop, Granada, Spain vol.~3 Citeseer.
\bibAnnoteFile{bergstra2011theano}

\bibitem[{Bergstra et~al.(2010)Bergstra, Breuleux, Bastien, Lamblin, Pascanu,
  Desjardins, Turian, Warde-Farley, and Bengio}]{bergstra2010theano}
Bergstra, J., O.~Breuleux, F.~Bastien, P.~Lamblin, R.~Pascanu, G.~Desjardins,
  J.~Turian, D.~Warde-Farley, and Y.~Bengio. 2010. Theano: a {CPU} and {GPU}
  math expression compiler. \emph{in} Proceedings of the Python for Scientific
  Computing Conference ({SciPy}) oral Presentation.
\bibAnnoteFile{bergstra2010theano}

\bibitem[{Bhattacharya et~al.(2016)Bhattacharya, Chakraborty, and
  Mallick}]{bhattacharya2016fast_sampling}
Bhattacharya, A., A.~Chakraborty, and B.~K. Mallick. 2016. Fast sampling with
  {G}aussian scale mixture priors in high-dimensional regression. Biometrika
  103:985--991.
\bibAnnoteFile{bhattacharya2016fast_sampling}

\bibitem[{Bletsa et~al.(2019)Bletsa, Suchard, Ji, Gryseels, Vrancken, Baele,
  Worobey, and Lemey}]{bletsa2019divergence}
Bletsa, M., M.~A. Suchard, X.~Ji, S.~Gryseels, B.~Vrancken, G.~Baele,
  M.~Worobey, and P.~Lemey. 2019. Divergence dating using mixed effects clock
  modelling: {An} application to {HIV}-1. Virus Evol 5:vez036.
\bibAnnoteFile{bletsa2019divergence}

\bibitem[{Bliss(1935)}]{bliss}
Bliss, C.~I. 1935. The comparison of dosage-mortality data. Annals of Applied
  Biology 22:307--333 (Fisher introduces his scoring method in appendix.).
\bibAnnoteFile{bliss}

\bibitem[{Bloom et~al.(2017)Bloom, Black, and Rappuoli}]{bloom2017}
Bloom, D.~E., S.~Black, and R.~Rappuoli. 2017. Emerging infectious diseases: a
  proactive approach. Proceedings of the {National Academy of Sciences}
  114:4055--4059.
\bibAnnoteFile{bloom2017}

\bibitem[{Bouchard-C{\^o}t{\'e} et~al.(2018)Bouchard-C{\^o}t{\'e}, Vollmer, and
  Doucet}]{bouchard2018bouncy}
Bouchard-C{\^o}t{\'e}, A., S.~J. Vollmer, and A.~Doucet. 2018. The bouncy
  particle sampler: A nonreversible rejection-free {M}arkov chain {Monte Carlo}
  method. Journal of the American Statistical Association 113:855--867.
\bibAnnoteFile{bouchard2018bouncy}

\bibitem[{Boyd et~al.(2004)Boyd, Boyd, and Vandenberghe}]{boyd2004convex}
Boyd, S., S.~P. Boyd, and L.~Vandenberghe. 2004. Convex optimization. Cambridge
  university press.
\bibAnnoteFile{boyd2004convex}

\bibitem[{Boyer et~al.(1998)Boyer, Brassard, H{\o}yer, and
  Tapp}]{boyer1998tight}
Boyer, M., G.~Brassard, P.~H{\o}yer, and A.~Tapp. 1998. Tight bounds on quantum
  searching. Fortschritte der Physik: Progress of Physics 46:493--505.
\bibAnnoteFile{boyer1998tight}

\bibitem[{Brooks et~al.(2011)Brooks, Gelman, Jones, and
  Meng}]{brooks2011handbook}
Brooks, S., A.~Gelman, G.~Jones, and X.-L. Meng. 2011. Handbook of {M}arkov
  chain {M}onte {C}arlo. CRC press.
\bibAnnoteFile{brooks2011handbook}

\bibitem[{Carvalho et~al.(2010)Carvalho, Polson, and
  Scott}]{carvalho2010horseshoe}
Carvalho, C.~M., N.~G. Polson, and J.~G. Scott. 2010. The horseshoe estimator
  for sparse signals. Biometrika 97:465--480.
\bibAnnoteFile{carvalho2010horseshoe}

\bibitem[{Chavan et~al.(2014)Chavan, Phursule et~al.}]{chavan2014survey}
Chavan, V., R.~N. Phursule, et~al. 2014. Survey paper on big data. Int. J.
  Comput. Sci. Inf. Technol 5:7932--7939.
\bibAnnoteFile{chavan2014survey}

\bibitem[{Cipra(2000)}]{cipra2000top10algo}
Cipra, B.~A. 2000. The best of the 20th century: Editors name top 10
  algorithms. SIAM news 33:1--2.
\bibAnnoteFile{cipra2000top10algo}

\bibitem[{Cleveland(2001)}]{cleveland2001data}
Cleveland, W.~S. 2001. Data science: an action plan for expanding the technical
  areas of the field of statistics. International statistical review 69:21--26.
\bibAnnoteFile{cleveland2001data}

\bibitem[{{COPSS}(2020)}]{copss2019}
{COPSS}. 2020. \emph{Committee of {P}residents of {S}tatistical {Societies}
  (Website)}. \url{https://community.amstat.org/copss/awards/winners}
  [Accessed: August 31, 2020].
\bibAnnoteFile{copss2019}

\bibitem[{Dagum and Menon(1998)}]{dagum1998openmp}
Dagum, L. and R.~Menon. 1998. Openmp: an industry standard api for
  shared-memory programming. IEEE computational science and engineering
  5:46--55.
\bibAnnoteFile{dagum1998openmp}

\bibitem[{Davenport and Patil(2012)}]{davenport2012data}
Davenport, T.~H. and D.~Patil. 2012. Data scientist. Harvard business review
  90:70--76.
\bibAnnoteFile{davenport2012data}

\bibitem[{Davis(1991)}]{davis1991handbook}
Davis, L. 1991. Handbook of genetic algorithms .
\bibAnnoteFile{davis1991handbook}

\bibitem[{Dempster et~al.(1977)Dempster, Laird, and
  Rubin}]{dempster1977maximum}
Dempster, A.~P., N.~M. Laird, and D.~B. Rubin. 1977. Maximum likelihood from
  incomplete data via the {EM} algorithm. Journal of the Royal Statistical
  Society: Series B (Methodological) 39:1--22.
\bibAnnoteFile{dempster1977maximum}

\bibitem[{Dinh et~al.(2017)Dinh, Bilge, Zhang, and
  Matsen~IV}]{dinh2017probabilistic-hmc}
Dinh, V., A.~Bilge, C.~Zhang, and F.~A. Matsen~IV. 2017. Probabilistic path
  {H}amiltonian {M}onte {C}arlo. Pages~1009--1018 \emph{in} Proceedings of the
  34th International Conference on Machine Learning vol.~70.
\bibAnnoteFile{dinh2017probabilistic-hmc}

\bibitem[{Dongarra et~al.(2016)Dongarra, Heroux, and
  Luszczek}]{dongarra2016hpc_connjugate_gradient}
Dongarra, J., M.~A. Heroux, and P.~Luszczek. 2016. High-performance
  conjugate-gradient benchmark: A new metric for ranking high-performance
  computing systems. The International Journal of High Performance Computing
  Applications 30:3--10.
\bibAnnoteFile{dongarra2016hpc_connjugate_gradient}

\bibitem[{Donoho(2017)}]{donoho201750}
Donoho, D. 2017. 50 years of data science. Journal of {Computational and
  Graphical Statistics} 26:745--766.
\bibAnnoteFile{donoho201750}

\bibitem[{Dudas et~al.(2017)Dudas, Carvalho, Bedford, Tatem, Baele, Faria,
  Park, Ladner, Arias, Asogun et~al.}]{dudas2017virus}
Dudas, G., L.~M. Carvalho, T.~Bedford, A.~J. Tatem, G.~Baele, N.~R. Faria,
  D.~J. Park, J.~T. Ladner, A.~Arias, D.~Asogun, et~al. 2017. Virus genomes
  reveal factors that spread and sustained the ebola epidemic. Nature
  544:309--315.
\bibAnnoteFile{dudas2017virus}

\bibitem[{Durr and Hoyer(1996)}]{durr1996quantum}
Durr, C. and P.~Hoyer. 1996. A quantum algorithm for finding the minimum. arXiv
  preprint quant-ph/9607014 .
\bibAnnoteFile{durr1996quantum}

\bibitem[{Efron(1992)}]{efron1992bootstrap}
Efron, B. 1992. Bootstrap methods: another look at the jackknife.
  Pages~569--593 \emph{in} Breakthroughs in statistics. Springer.
\bibAnnoteFile{efron1992bootstrap}

\bibitem[{Efron and Tibshirani(1994)}]{efron1994introduction}
Efron, B. and R.~J. Tibshirani. 1994. An introduction to the bootstrap. CRC
  press.
\bibAnnoteFile{efron1994introduction}

\bibitem[{Elbe and Buckland-Merrett(2017)}]{elbe2017data}
Elbe, S. and G.~Buckland-Merrett. 2017. Data, disease and diplomacy: Gisaid's
  innovative contribution to global health. Global Challenges 1:33--46.
\bibAnnoteFile{elbe2017data}

\bibitem[{Fisher(1922)}]{fisher1922mathematical}
Fisher, R.~A. 1922. On the mathematical foundations of theoretical statistics.
  Philosophical Transactions of the Royal Society of London. Series A,
  Containing Papers of a Mathematical or Physical Character 222:309--368.
\bibAnnoteFile{fisher1922mathematical}

\bibitem[{Fisher(1960)}]{fisher1960design}
Fisher, R.~A. 1960. The design of experiments. {(Especially Section 21.)}.
\bibAnnoteFile{fisher1960design}

\bibitem[{Fisher(1992)}]{fisher1992statistical}
Fisher, R.~A. 1992. Statistical methods for research workers. \emph{in}
  Breakthroughs in statistics. Springer {(Especially Section 21.02)}.
\bibAnnoteFile{fisher1992statistical}

\bibitem[{Friedman et~al.(2010)Friedman, Hastie, and
  Tibshirani}]{friedman2010coord-descent}
Friedman, J., T.~Hastie, and R.~Tibshirani. 2010. Regularization paths for
  generalized linear models via coordinate descent. Journal of Statistical
  Software 33:1.
\bibAnnoteFile{friedman2010coord-descent}

\bibitem[{Gelfand and Smith(1990)}]{gelfand1990sampling}
Gelfand, A.~E. and A.~F. Smith. 1990. Sampling-based approaches to calculating
  marginal densities. Journal of the American statistical association
  85:398--409.
\bibAnnoteFile{gelfand1990sampling}

\bibitem[{Gelman(2014)}]{gelman2014petascale_grant_report}
Gelman, A. 2014. Petascale hierarchical modeling via parallel execution. U.S.
  Department of Energy. Report No: DE-SC0002099 .
\bibAnnoteFile{gelman2014petascale_grant_report}

\bibitem[{Gelman et~al.(2013)Gelman, Carlin, Stern, Dunson, Vehtari, and
  Rubin}]{gelman2013bayesian}
Gelman, A., J.~B. Carlin, H.~S. Stern, D.~B. Dunson, A.~Vehtari, and D.~B.
  Rubin. 2013. Bayesian data analysis. CRC press.
\bibAnnoteFile{gelman2013bayesian}

\bibitem[{Gelman et~al.(1996)Gelman, Roberts, Gilks
  et~al.}]{gelman1996efficient}
Gelman, A., G.~O. Roberts, W.~R. Gilks, et~al. 1996. Efficient metropolis
  jumping rules. Bayesian statistics 5:42.
\bibAnnoteFile{gelman1996efficient}

\bibitem[{Geman and Geman(1984)}]{geman1984stochastic}
Geman, S. and D.~Geman. 1984. Stochastic relaxation, gibbs distributions, and
  the bayesian restoration of images. IEEE Transactions on pattern analysis and
  machine intelligence Pages~721--741.
\bibAnnoteFile{geman1984stochastic}

\bibitem[{Gentle et~al.(2012)Gentle, H{\"a}rdle, and
  Mori}]{gentle2012computational}
Gentle, J.~E., W.~K. H{\"a}rdle, and Y.~Mori. 2012. How computational
  statistics became the backbone of modern data science. Pages~3--16 \emph{in}
  Handbook of Computational Statistics. Springer.
\bibAnnoteFile{gentle2012computational}

\bibitem[{George and McCulloch(1997)}]{george1997approaches}
George, E.~I. and R.~E. McCulloch. 1997. Approaches for bayesian variable
  selection. Statistica sinica Pages~339--373.
\bibAnnoteFile{george1997approaches}

\bibitem[{Geyer(1991)}]{geyer1991markov}
Geyer, C.~J. 1991. Markov chain {Monte Carlo} maximum likelihood .
\bibAnnoteFile{geyer1991markov}

\bibitem[{Ghosh and Clyde(2011)}]{ghosh2011design_orthogonalization}
Ghosh, J. and M.~A. Clyde. 2011. {R}ao--{B}lackwellization for {B}ayesian
  variable selection and model averaging in linear and binary regression: a
  novel data augmentation approach. Journal of the American Statistical
  Association 106:1041--1052.
\bibAnnoteFile{ghosh2011design_orthogonalization}

\bibitem[{Golub and Van~Loan(2012)}]{golub2012matrix_computation}
Golub, G.~H. and C.~F. Van~Loan. 2012. Matrix computations vol.~3. Johns
  Hopkins University Press.
\bibAnnoteFile{golub2012matrix_computation}

\bibitem[{{Google Trends}(2020)}]{trends}
{Google Trends}. 2020. Data source: Google trends.
\bibAnnoteFile{trends}

\bibitem[{Grover(1996)}]{grover1996fast}
Grover, L.~K. 1996. A fast quantum mechanical algorithm for database search.
  Pages~212--219 \emph{in} Proceedings of the twenty-eighth annual ACM
  symposium on Theory of computing.
\bibAnnoteFile{grover1996fast}

\bibitem[{Harrow et~al.(2009)Harrow, Hassidim, and Lloyd}]{harrow2009quantum}
Harrow, A.~W., A.~Hassidim, and S.~Lloyd. 2009. Quantum algorithm for linear
  systems of equations. Physical review letters 103:150502.
\bibAnnoteFile{harrow2009quantum}

\bibitem[{Hastie et~al.(2015)Hastie, Tibshirani, and
  Wainwright}]{hastie2015learning-with-sparsity}
Hastie, T., R.~Tibshirani, and M.~Wainwright. 2015. Statistical learning with
  sparsity: the lasso and generalizations. CRC press.
\bibAnnoteFile{hastie2015learning-with-sparsity}

\bibitem[{Hastings(1970)}]{hastings1970monte}
Hastings, W.~K. 1970. Monte {C}arlo sampling methods using {M}arkov chains and
  their applications .
\bibAnnoteFile{hastings1970monte}

\bibitem[{Hestenes and Stiefel(1952)}]{hestenes1952cg}
Hestenes, M.~R. and E.~Stiefel. 1952. Methods of conjugate gradients for
  solving linear systems. Journal of Research of the National Bureau of
  Standards 49.
\bibAnnoteFile{hestenes1952cg}

\bibitem[{Hocking and Leslie(1967)}]{hocking1967selection}
Hocking, R.~R. and R.~Leslie. 1967. Selection of the best subset in regression
  analysis. Technometrics 9:531--540.
\bibAnnoteFile{hocking1967selection}

\bibitem[{Hoffman and Gelman(2014)}]{hoffman2014nuts}
Hoffman, M.~D. and A.~Gelman. 2014. The no-{U}-turn sampler: adaptively setting
  path lengths in {H}amiltonian {M}onte {C}arlo. Journal of Machine Learning
  Research 15:1593--1623.
\bibAnnoteFile{hoffman2014nuts}

\bibitem[{Holbrook et~al.(2017)Holbrook, Vandenberg-Rodes, Fortin, and
  Shahbaba}]{holbrook2017bayesian}
Holbrook, A., A.~Vandenberg-Rodes, N.~Fortin, and B.~Shahbaba. 2017. A bayesian
  supervised dual-dimensionality reduction model for simultaneous decoding of
  lfp and spike train signals. Stat 6:53--67.
\bibAnnoteFile{holbrook2017bayesian}

\bibitem[{Holbrook(2021)}]{holbrook2021quantum}
Holbrook, A.~J. 2021. A quantum parallel markov chain monte carlo. arXiv
  preprint arXiv:2112.00212 .
\bibAnnoteFile{holbrook2021quantum}

\bibitem[{Holbrook et~al.(2020{\natexlab{a}})Holbrook, Lemey, Baele, Dellicour,
  Brockmann, Rambaut, and Suchard}]{holbrook2020massive}
Holbrook, A.~J., P.~Lemey, G.~Baele, S.~Dellicour, D.~Brockmann, A.~Rambaut,
  and M.~A. Suchard. 2020{\natexlab{a}}. Massive parallelization boosts big
  bayesian multidimensional scaling. Journal of Computational and Graphical
  Statistics Pages~1--34.
\bibAnnoteFile{holbrook2020massive}

\bibitem[{Holbrook et~al.(2020{\natexlab{b}})Holbrook, Loeffler, Flaxman, and
  Suchard}]{holbrook2020scalable}
Holbrook, A.~J., C.~E. Loeffler, S.~R. Flaxman, and M.~A. Suchard.
  2020{\natexlab{b}}. Scalable bayesian inference for self-excitatory
  stochastic processes applied to big american gunfire data. arXiv preprint
  arXiv:2005.10123 .
\bibAnnoteFile{holbrook2020scalable}

\bibitem[{Hunter and Lange(2004)}]{hunter2004tutorial}
Hunter, D.~R. and K.~Lange. 2004. A tutorial on mm algorithms. The American
  Statistician 58:30--37.
\bibAnnoteFile{hunter2004tutorial}

\bibitem[{Ji et~al.(2020)Ji, Zhang, Holbrook, Nishimura, Baele, Rambaut, Lemey,
  and Suchard}]{ji2020gradients}
Ji, X., Z.~Zhang, A.~Holbrook, A.~Nishimura, G.~Baele, A.~Rambaut, P.~Lemey,
  and M.~A. Suchard. 2020. Gradients do grow on trees: a linear-time
  {O(N)}-dimensional gradient for statistical phylogenetics. Molecular {Biology
  and Evolution} .
\bibAnnoteFile{ji2020gradients}

\bibitem[{Jordan et~al.(1999)Jordan, Ghahramani, Jaakkola, and
  Saul}]{jordan1999variational}
Jordan, M.~I., Z.~Ghahramani, T.~S. Jaakkola, and L.~K. Saul. 1999. An
  introduction to variational methods for graphical models. Machine learning
  37:183--233.
\bibAnnoteFile{jordan1999variational}

\bibitem[{Jordan(2005)}]{jordan2005fast}
Jordan, S.~P. 2005. Fast quantum algorithm for numerical gradient estimation.
  Physical review letters 95:050501.
\bibAnnoteFile{jordan2005fast}

\bibitem[{Kahle and Wickham(2013)}]{kahle2013ggmap}
Kahle, D. and H.~Wickham. 2013. ggmap: Spatial visualization with ggplot2. The
  {R} {J}ournal 5:144--161.
\bibAnnoteFile{kahle2013ggmap}

\bibitem[{Kennedy and Eberhart(1995)}]{kennedy1995particle}
Kennedy, J. and R.~Eberhart. 1995. Particle swarm optimization.
  Pages~1942--1948 \emph{in} Proceedings of ICNN'95-International Conference on
  Neural Networks vol.~4 IEEE.
\bibAnnoteFile{kennedy1995particle}

\bibitem[{Lan et~al.(2014)Lan, Streets, and Shahbaba}]{lan2014wormhole}
Lan, S., J.~Streets, and B.~Shahbaba. 2014. Wormhole {H}amiltonian {M}onte
  {C}arlo. \emph{in} Twenty-Eighth AAAI Conference on Artificial Intelligence.
\bibAnnoteFile{lan2014wormhole}

\bibitem[{Lanczos(1952)}]{lanczos1952iterative}
Lanczos, C. 1952. Solution of systems of linear equations by minimized
  iterations. Journal of Research of the National Bureau of Standards
  49:33--53.
\bibAnnoteFile{lanczos1952iterative}

\bibitem[{Livingstone et~al.(2019)Livingstone, Faulkner, and
  Roberts}]{livingstone2019kinetic_energy_choice}
Livingstone, S., M.~F. Faulkner, and G.~O. Roberts. 2019. Kinetic energy choice
  in {H}amiltonian/hybrid {M}onte {C}arlo. Biometrika 106:303--319.
\bibAnnoteFile{livingstone2019kinetic_energy_choice}

\bibitem[{Livingstone and Zanella(2019)}]{livingstone2019robustness}
Livingstone, S. and G.~Zanella. 2019. On the robustness of gradient-based mcmc
  algorithms. ar{X}iv:1908.11812 .
\bibAnnoteFile{livingstone2019robustness}

\bibitem[{Lunn et~al.(2009)Lunn, Spiegelhalter, Thomas, and
  Best}]{lunn2009bugs}
Lunn, D., D.~Spiegelhalter, A.~Thomas, and N.~Best. 2009. The {BUGS} project:
  Evolution, critique and future directions. Statistics in Medicine
  28:3049--3067.
\bibAnnoteFile{lunn2009bugs}

\bibitem[{Madigan and Raftery(1994)}]{madigan1994model}
Madigan, D. and A.~E. Raftery. 1994. Model selection and accounting for model
  uncertainty in graphical models using occam's window. Journal of the American
  Statistical Association 89:1535--1546.
\bibAnnoteFile{madigan1994model}

\bibitem[{Mangoubi et~al.(2018)Mangoubi, Pillai, and
  Smith}]{mangoubi2018hmc_mixing_on_multimodal}
Mangoubi, O., N.~S. Pillai, and A.~Smith. 2018. Does {H}amiltonian {M}onte
  {C}arlo mix faster than a random walk on multimodal densities?
  arXiv:1808.03230 .
\bibAnnoteFile{mangoubi2018hmc_mixing_on_multimodal}

\bibitem[{McCullagh and Nelder(1989)}]{mccullagh1989generalized}
McCullagh, P. and J.~Nelder. 1989. Generalized linear models., 2nd edn.(chapman
  and hall: London). Standard book on generalized linear models .
\bibAnnoteFile{mccullagh1989generalized}

\bibitem[{Metropolis et~al.(1953)Metropolis, Rosenbluth, Rosenbluth, Teller,
  and Teller}]{metropolis1953equation}
Metropolis, N., A.~W. Rosenbluth, M.~N. Rosenbluth, A.~H. Teller, and
  E.~Teller. 1953. Equation of state calculations by fast computing machines.
  The journal of chemical physics 21:1087--1092.
\bibAnnoteFile{metropolis1953equation}

\bibitem[{Mitchell and Beauchamp(1988)}]{mitchell1988bayesian}
Mitchell, T.~J. and J.~J. Beauchamp. 1988. Bayesian variable selection in
  linear regression. Journal of the american statistical association
  83:1023--1032.
\bibAnnoteFile{mitchell1988bayesian}

\bibitem[{Monnahan et~al.(2017)Monnahan, Thorson, and
  Branch}]{monnahan2017faster}
Monnahan, C.~C., J.~T. Thorson, and T.~A. Branch. 2017. Faster estimation of
  {B}ayesian models in ecology using {Hamiltonian Monte Carlo}. Methods in
  {Ecology and Evolution} 8:339--348.
\bibAnnoteFile{monnahan2017faster}

\bibitem[{Murty and Kabadi(1985)}]{murty1985some}
Murty, K.~G. and S.~N. Kabadi. 1985. Some np-complete problems in quadratic and
  nonlinear programming. Tech. rep.
\bibAnnoteFile{murty1985some}

\bibitem[{Neal(1996)}]{neal1996bayesian_neuralnet}
Neal, R.~M. 1996. Bayesian Learning for Neural Networks. Springer-Verlag.
\bibAnnoteFile{neal1996bayesian_neuralnet}

\bibitem[{Neal(2011)}]{neal2011mcmc}
Neal, R.~M. 2011. {MCMC} using {H}amiltonian {D}ynamics. \emph{in} Handbook of
  {M}arkov chain {M}onte {C}arlo. CRC Press.
\bibAnnoteFile{neal2011mcmc}

\bibitem[{Nielsen and Chuang(2002)}]{nielsen2002quantum}
Nielsen, M.~A. and I.~Chuang. 2002. Quantum computation and quantum
  information.
\bibAnnoteFile{nielsen2002quantum}

\bibitem[{Nishimura and Dunson(2016)}]{nishimura2016geometrically}
Nishimura, A. and D.~Dunson. 2016. Geometrically tempered {H}amiltonian {M}onte
  {C}arlo. arXiv preprint arXiv:1604.00872 .
\bibAnnoteFile{nishimura2016geometrically}

\bibitem[{Nishimura et~al.(2020)Nishimura, Dunson, and
  Lu}]{nishimura2020discontinuous}
Nishimura, A., D.~B. Dunson, and J.~Lu. 2020. Discontinuous {Hamiltonian Monte
  Carlo} for discrete parameters and discontinuous likelihoods. Biometrika
  107:365--380.
\bibAnnoteFile{nishimura2020discontinuous}

\bibitem[{Nishimura and Suchard(2018)}]{nishimura2018cg_accelerated_gibbs}
Nishimura, A. and M.~A. Suchard. 2018. Prior-preconditioned conjugate gradient
  for accelerated gibbs sampling in" large n \& large p" sparse bayesian
  logistic regression models. ar{X}iv:1810.12437 .
\bibAnnoteFile{nishimura2018cg_accelerated_gibbs}

\bibitem[{Nott and
  Kohn(2005)}]{nott2005adaptive_sampling_for_variable_selection}
Nott, D.~J. and R.~Kohn. 2005. Adaptive sampling for {B}ayesian variable
  selection. Biometrika 92:747--763.
\bibAnnoteFile{nott2005adaptive_sampling_for_variable_selection}

\bibitem[{Nunes et~al.(2014)Nunes, Palacios, Faria, Sousa~Jr, Pantoja,
  Rodrigues, Carvalho, Medeiros, Savji, Baele et~al.}]{nunes2014air}
Nunes, M.~R., G.~Palacios, N.~R. Faria, E.~C. Sousa~Jr, J.~A. Pantoja, S.~G.
  Rodrigues, V.~L. Carvalho, D.~B. Medeiros, N.~Savji, G.~Baele, et~al. 2014.
  Air travel is associated with intracontinental spread of dengue virus
  serotypes 1--3 in {Brazil}. {PLoS Neglected Tropical Diseases} 8:e2769.
\bibAnnoteFile{nunes2014air}

\bibitem[{Passos et~al.(2019)Passos, Mwangi, and
  Kapczinski}]{passos2019personalized_psychiatry}
Passos, I.~C., B.~Mwangi, and F.~Kapczinski. 2019. Personalized psychiatry: big
  data analytics in mental health. Springer.
\bibAnnoteFile{passos2019personalized_psychiatry}

\bibitem[{Polson and Scott(2010)}]{polson2010global_local}
Polson, N.~G. and J.~G. Scott. 2010. Shrink globally, act locally: Sparse
  {B}ayesian regularization and prediction. Bayesian Statistics 9:501--538.
\bibAnnoteFile{polson2010global_local}

\bibitem[{Polson et~al.(2013)Polson, Scott, and Windle}]{polson2013polya_gamma}
Polson, N.~G., J.~G. Scott, and J.~Windle. 2013. {B}ayesian inference for
  logistic models using {P}{\'o}lya--{G}amma latent variables. Journal of the
  American Statistical Association 108:1339--1349.
\bibAnnoteFile{polson2013polya_gamma}

\bibitem[{Pybus et~al.(2012)Pybus, Suchard, Lemey, Bernardin, Rambaut,
  Crawford, Gray, Arinaminpathy, Stramer, Busch et~al.}]{Pybus2012}
Pybus, O.~G., M.~A. Suchard, P.~Lemey, F.~J. Bernardin, A.~Rambaut, F.~W.
  Crawford, R.~R. Gray, N.~Arinaminpathy, S.~L. Stramer, M.~P. Busch, et~al.
  2012. Unifying the spatial epidemiology and molecular evolution of emerging
  epidemics. Proceedings of the {National Academy of Sciences}
  109:15066--15071.
\bibAnnoteFile{Pybus2012}

\bibitem[{Pybus et~al.(2015)Pybus, Tatem, and Lemey}]{Pybus2015}
Pybus, O.~G., A.~J. Tatem, and P.~Lemey. 2015. Virus evolution and transmission
  in an ever more connected world. Proceedings of the {Royal Society B:
  Biological Sciences} 282:20142878.
\bibAnnoteFile{Pybus2015}

\bibitem[{Ranganath et~al.(2014)Ranganath, Gerrish, and
  Blei}]{ranganath2014black}
Ranganath, R., S.~Gerrish, and D.~M. Blei. 2014. Black box variational
  inference. \emph{in} Proceedings of the Seventeenth International Conference
  on Artificial Intelligence and Statistics.
\bibAnnoteFile{ranganath2014black}

\bibitem[{Rue and Held(2005)}]{rue2005gmrf}
Rue, H. and L.~Held. 2005. Gaussian {M}arkov random fields: theory and
  applications. CRC press.
\bibAnnoteFile{rue2005gmrf}

\bibitem[{Rumelhart et~al.(1986)Rumelhart, Hinton, and
  Williams}]{rumelhart1986learning}
Rumelhart, D.~E., G.~E. Hinton, and R.~J. Williams. 1986. Learning
  representations by back-propagating errors. nature 323:533--536.
\bibAnnoteFile{rumelhart1986learning}

\bibitem[{Seber and Lee(2012)}]{seber2012linear}
Seber, G.~A. and A.~J. Lee. 2012. Linear regression analysis vol. 329. John
  Wiley \& Sons.
\bibAnnoteFile{seber2012linear}

\bibitem[{{Stan Development Team}(2018)}]{stan2018}
{Stan Development Team}. 2018. Stan Modeling Language Users Guide and Reference
  Manual, Version 2.18.0.
\bibAnnoteFile{stan2018}

\bibitem[{Suchard et~al.(2018)Suchard, Lemey, Baele, Ayres, Drummond, and
  Rambaut}]{suchard2018bayesian}
Suchard, M.~A., P.~Lemey, G.~Baele, D.~L. Ayres, A.~J. Drummond, and
  A.~Rambaut. 2018. Bayesian phylogenetic and phylodynamic data integration
  using {BEAST} 1.10. Virus {E}volution 4:vey016.
\bibAnnoteFile{suchard2018bayesian}

\bibitem[{Suchard et~al.(2019)Suchard, Schuemie, Krumholz, You, Chen, Pratt,
  Reich, Duke, Madigan, Hripcsak et~al.}]{suchard2019legend}
Suchard, M.~A., M.~J. Schuemie, H.~M. Krumholz, S.~C. You, R.~Chen, N.~Pratt,
  C.~G. Reich, J.~Duke, D.~Madigan, G.~Hripcsak, et~al. 2019. Comprehensive
  comparative effectiveness and safety of first-line antihypertensive drug
  classes: a systematic, multinational, large-scale analysis. The Lancet
  394:1816--1826.
\bibAnnoteFile{suchard2019legend}

\bibitem[{Svensson et~al.(2019)Svensson, da~Veiga~Beltrame, and
  Pachter}]{svensson2019curated}
Svensson, V., E.~da~Veiga~Beltrame, and L.~Pachter. 2019. A curated database
  reveals trends in single-cell transcriptomics. BioRxiv Page~742304.
\bibAnnoteFile{svensson2019curated}

\bibitem[{Tibshirani(1996)}]{tibshirani1996regression}
Tibshirani, R. 1996. Regression shrinkage and selection via the lasso. Journal
  of the Royal Statistical Society: Series B (Methodological) 58:267--288.
\bibAnnoteFile{tibshirani1996regression}

\bibitem[{Tierney(1994)}]{tierney1994markov}
Tierney, L. 1994. Markov chains for exploring posterior distributions. the
  Annals of Statistics Pages~1701--1728.
\bibAnnoteFile{tierney1994markov}

\bibitem[{Tjelmeland and Hegstad(2001)}]{tjelmeland2001mode}
Tjelmeland, H. and B.~K. Hegstad. 2001. Mode jumping proposals in mcmc.
  Scandinavian journal of statistics 28:205--223.
\bibAnnoteFile{tjelmeland2001mode}

\bibitem[{Trefethen and Bau(1997)}]{trefethen1997numerical_linalg}
Trefethen, L.~N. and D.~Bau. 1997. Numerical Linear Algebra. Society for
  Industrial and Applied Mathematics.
\bibAnnoteFile{trefethen1997numerical_linalg}

\bibitem[{Van~der Vorst(2003)}]{vorst2003iterative}
Van~der Vorst, H.~A. 2003. Iterative {K}rylov Methods for Large Linear Systems
  vol.~13. Cambridge University Press.
\bibAnnoteFile{vorst2003iterative}

\bibitem[{Van~Dyk and Meng(2001)}]{van2001art}
Van~Dyk, D.~A. and X.-L. Meng. 2001. The art of data augmentation. Journal of
  Computational and Graphical Statistics 10:1--50.
\bibAnnoteFile{van2001art}

\bibitem[{Wald and Wolfowitz(1944)}]{wald1944statistical}
Wald, A. and J.~Wolfowitz. 1944. Statistical tests based on permutations of the
  observations. The Annals of Mathematical Statistics 15:358--372.
\bibAnnoteFile{wald1944statistical}

\bibitem[{Warne et~al.(2019)Warne, Sisson, and
  Drovandi}]{warne2019acceleration}
Warne, D.~J., S.~A. Sisson, and C.~Drovandi. 2019. Acceleration of expensive
  computations in bayesian statistics using vector operations. arXiv preprint
  arXiv:1902.09046 .
\bibAnnoteFile{warne2019acceleration}

\bibitem[{Wei and Tanner(1990)}]{wei1990monte}
Wei, G.~C. and M.~A. Tanner. 1990. A {Monte Carlo} implementation of the {EM}
  algorithm and the poor man's data augmentation algorithms. Journal of the
  American statistical Association 85:699--704.
\bibAnnoteFile{wei1990monte}

\bibitem[{Wickham(2016)}]{wickham2016ggplot2}
Wickham, H. 2016. ggplot2: elegant graphics for data analysis. springer.
\bibAnnoteFile{wickham2016ggplot2}

\bibitem[{Wickham et~al.(2007)}]{wickham2007reshaping}
Wickham, H. et~al. 2007. Reshaping data with the reshape package. Journal of
  {Statistical Software} 21:1--20.
\bibAnnoteFile{wickham2007reshaping}

\bibitem[{Wickham et~al.(2011)}]{wickham2011split}
Wickham, H. et~al. 2011. The split-apply-combine strategy for data analysis.
  Journal of {Statistical Software} 40:1--29.
\bibAnnoteFile{wickham2011split}

\bibitem[{Wickham et~al.(2014)}]{wickham2014tidy}
Wickham, H. et~al. 2014. Tidy data. Journal of {Statistical Software} 59:1--23.
\bibAnnoteFile{wickham2014tidy}

\bibitem[{Williams and Rasmussen(1996)}]{williams1996gaussian}
Williams, C.~K. and C.~E. Rasmussen. 1996. Gaussian processes for regression.
  Pages~514--520 \emph{in} Advances in neural information processing systems.
\bibAnnoteFile{williams1996gaussian}

\bibitem[{Williams and Rasmussen(2006)}]{williams2006gaussian}
Williams, C.~K. and C.~E. Rasmussen. 2006. Gaussian processes for machine
  learning vol.~2. MIT press Cambridge, MA.
\bibAnnoteFile{williams2006gaussian}

\bibitem[{{\disambiguate{Zhang, L.}{L.}{Zhang}}
  et~al.(2019){\disambiguate{Zhang, L.}{L.}{Zhang}}, Datta, and
  Banerjee}]{zhang2019practical-scalable-computing}
{\disambiguate{Zhang, L.}{L.}{Zhang}}, A.~Datta, and S.~Banerjee. 2019.
  Practical {B}ayesian modeling and inference for massive spatial data sets on
  modest computing environments. Statistical Analysis and Data Mining: The ASA
  Data Science Journal 12:197--209.
\bibAnnoteFile{zhang2019practical-scalable-computing}

\bibitem[{{\disambiguate{Zhang, Z.}{Z.}{Zhang}}
  et~al.(2020){\disambiguate{Zhang, Z.}{Z.}{Zhang}}, Nishimura, Bastide, Ji,
  Payne, Goulder, Lemey, and Suchard}]{zhang2019large}
{\disambiguate{Zhang, Z.}{Z.}{Zhang}}, A.~Nishimura, P.~Bastide, X.~Ji, R.~P.
  Payne, P.~Goulder, P.~Lemey, and M.~A. Suchard. 2020. Large-scale inference
  of correlation among mixed-type biological traits with phylogenetic
  multivariate probit models. {Annals of Applied S}tatistics. In press.
\bibAnnoteFile{zhang2019large}

\end{thebibliography}

\end{document}